\documentclass[aps,prb,preprint,,superscriptaddress,floatfix,showpacs]{revtex4-1}
\usepackage{hyperref}
\usepackage{graphicx}
\usepackage{amsfonts}
\usepackage{amssymb}
\usepackage{amsmath}
\usepackage{mathrsfs}

\usepackage{txfonts}
\usepackage{lipsum}
\usepackage{color}
\usepackage{wasysym}

\usepackage{letltxmacro}

\begin{document}
%\title{Berry's phase atomic interferometers in graphene.}
%\title{Wavefront dislocations in  Friedel oscillations reveal the topological features of band structures}
\title{Measuring the Berry phase of graphene from wavefront dislocations in Friedel oscillations}
%\title{Topological defects in Freidel oscillations reveal topological band structures.}
\author{C. Dutreix}
\email{email: clement.dutreix@u-bordeaux.fr; vincent.renard@cea.fr}
\affiliation{Universit\'e de Bordeaux, France and CNRS, LOMA, UMR 5798, Talence, F-33400, France}

\author{H. Gonz\'{a}lez-Herrero}
\affiliation{Departamento de F\'isica de la Materia Condensada, Universidad Aut\'onoma de Madrid, E-28049 Madrid, Spain.}
\affiliation{Condensed Matter Physics Center (IFIMAC), Universidad Aut\'onoma de Madrid, E-28049 Madrid, Spain.}

\author{I. Brihuega}
\affiliation{Departamento de F\'isica de la Materia Condensada, Universidad Aut\'onoma de Madrid, E-28049 Madrid, Spain.}
\affiliation{Condensed Matter Physics Center (IFIMAC), Universidad Aut\'onoma de Madrid, E-28049 Madrid, Spain.}
\affiliation{Instituto Nicol\'as Cabrera, Universidad Aut\'onoma de Madrid, E-28049 Madrid, Spain.}

\author{M. I. Katsnelson}
\affiliation{Radboud University, Institute for Molecules and Materials, Nijmegen, The Netherlands}

\author{C. Chapelier}
\affiliation{Univ. Grenoble Alpes, CEA, IRIG, PHELIQS, F-38000 Grenoble, France}

\author{V. T. Renard}
\email{email: clement.dutreix@u-bordeaux.fr; vincent.renard@cea.fr}
\affiliation{Univ. Grenoble Alpes, CEA, IRIG, PHELIQS, F-38000 Grenoble, France}

\iffalse
\begin{abstract}
\end{abstract}
\fi

\maketitle

{\bf
Electronic band structures dictate the mechanical, optical and electrical properties of crystalline solids. Their experimental determination is therefore of crucial importance for technological applications. While the spectral distribution in energy bands is routinely measured by various techniques,\cite{Solyom2009} it is more difficult to access the topological properties of band structures such as the Berry phase $\gamma$. It is usually thought that measuring the Berry phase requires applying external electromagnetic forces because these allow realizing the adiabatic transport on closed trajectories along which quantum mechanical wave-functions pick up the Berry phase.\cite{Berry:1984aa,Xiao:2010aa} In graphene, the anomalous quantum Hall effect results from the Berry phase $\gamma=\pi$ picked up by massless relativistic electrons along cyclotron orbits \cite{Novoselov:2005aa,Zhang:2005aa} and proves the existence of Dirac cones. Contradicting this belief, we demonstrate that the Berry phase of graphene can be measured in absence of any external magnetic field. We observe edge dislocations in the Friedel oscillations formed at hydrogen atoms chemisorbed on graphene.  Following Nye and Berry in describing these topological defects as phase singularities of complex fields,\cite{Nye:1974aa} we show that the number of additional wave-fronts in the dislocation is a real space measurement of the pseudo spin winding, i.e. graphene's Berry phase. Since the electronic dispersion can also be retrieved from Friedel oscillations,\cite{Crommie1993} our study establishes the electronic density as a powerful observable to determine both the dispersion relation and topological properties of wavefunctions. This could have profound consequences for the study of the band-structure topology of relativistic and gapped phases in solids.\cite{Xiao:2010aa}    
}

\begin{figure*}[]
	%\centering
	\includegraphics[width=\textwidth]{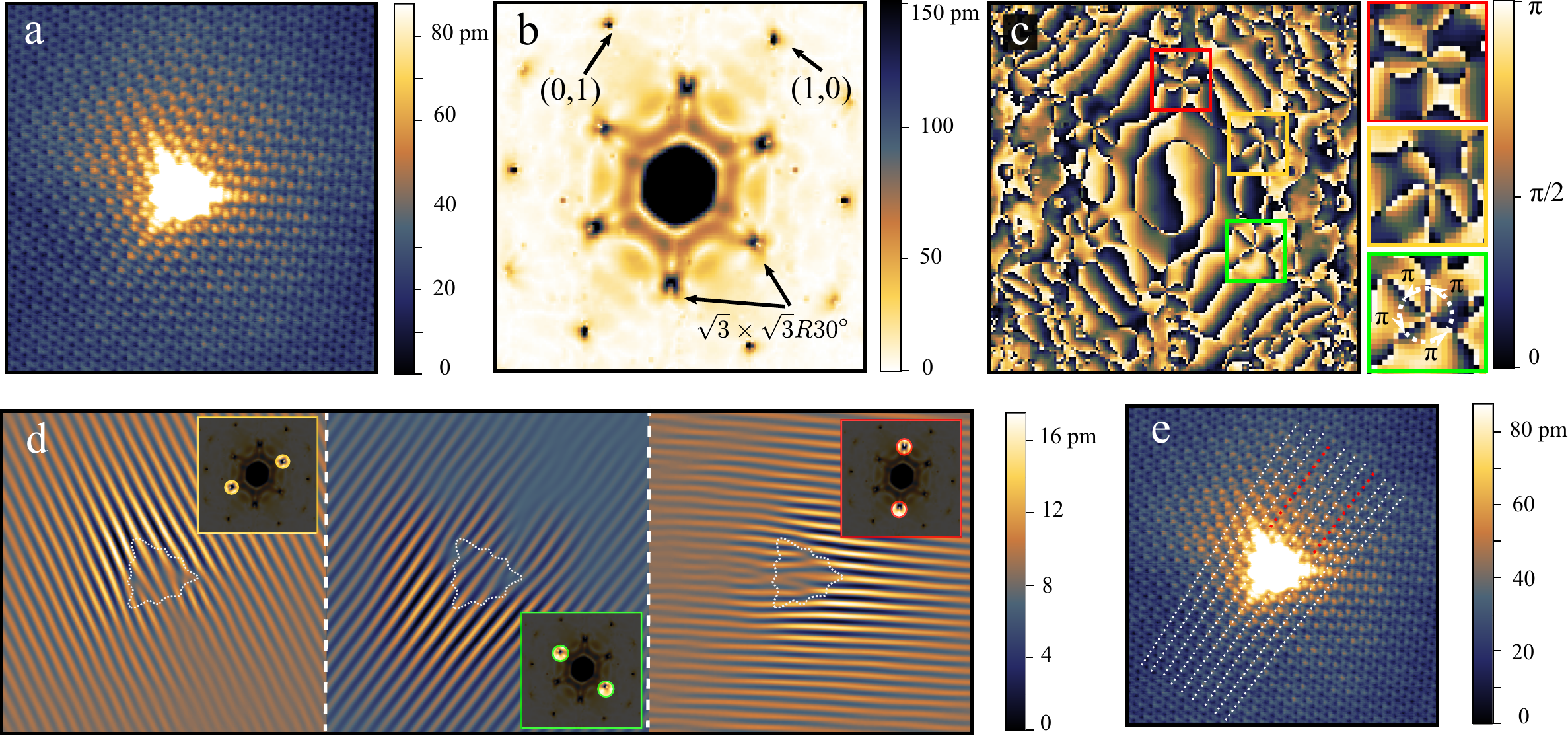}
	\caption{\textbf{Dislocations in Friedel oscillations near a H atom.} {\bf a}, Topography STM image of an hydrogen adatom at the surface of graphene. The image is 10 nm $\times$ 10 nm in size. The tunnelling bias is $V_b=0.4$~V and the tunnelling current is $i_t=45.5$~pA. \label{key} {\bf b}, Modulus of the Fast Fourier Transform (FFT) of the image in panel ({\bf a}). The points labeled (1,0) and (0,1) correspond to the atomic signal. {\bf c}, Phase of the Fast Fourier Transform (FFT) of the image in panel ({\bf a}). Zooms at the $\sqrt{3}\times\sqrt{3}R30^\circ$ points are presented on the right. The phase winds by $4\pi$ around each of these spots (The sharp boudary between bright and dark indicate a phase shift of $\pi$). The FFT images are 62.8 nm$^{-1}\times$62.8 nm$^{-1}$. {\bf d}, FFT filtered images of panel ({\bf a}) along the three directions of intervalley scattering  (The inset shows the filters applied in the Fourier space). {\bf e}, Raw images with dotted lines highlighting the wave front for one direction of intervalley scattering. The red dotted lines correspond to the additional wave-fronts. Similar result is obtained in the other directions (Supplementary information).    }
	\label{STM Dislocations}
\end{figure*}

Particle-wave duality manifests as an oscillatory structure in the static response of conduction electrons to impurities: Friedel oscillations.\cite{Friedel:1952aa} These appear in various contexts and can for example alter the conductance of two-dimensional electron gases\cite{Zala2001} or mediate long range interactions between magnetic impurities.\cite{Ruderman1954,Kasuya1956,Yosida1957} Since Friedel oscillations intrinsically result from the quantum interference of electronic waves, they necessarily carry information about the crystalline host materials which impose constrains on the possible wave-functions. For instance, the $2q_{\bf F}$ dependence of Friedel oscillations can be used to recover energy dispersion from a sequence of energy-resolved scanning tunnelling microscope (STM) images.\cite{Crommie1993} This was used to reconstruct the linear dispersion in graphene.\cite{Rutter:2007,Mallet:2012aa} Friedel oscillations were also used to demonstrate the existence of graphene's pseudospin.\cite{Brihuega:2008,Mallet:2012aa} However, the pseudospin winding, which is directly related to graphene's Berry phase and characterises the band-structure topology of massless relativistic electrons, has not been retrieved from such STM images.
Figure.~\ref{STM Dislocations}a shows an experimental STM image of a H atom chemisorbed on graphene (See Methods and Ref\,\onlinecite{Gonzalez-Herrero:2016aa} for experimental details). The $\sqrt{3}\times\sqrt{3}R30^\circ$ spots in the Fourier transform (Fig.~\ref{STM Dislocations}b and \ref{STM Dislocations}c) contain signatures of Friedel oscillations associated to elastic backscattering of Dirac electrons from a given valley \textbf{K} to a nearest-neighbour one \textbf{K'}.\cite{Rutter:2007,Mallet:2012aa,Brihuega:2008} Figure~\ref{STM Dislocations}d shows the corresponding oscillation in the real space after  Fourier filtering the signal for each direction of intervalley backscattering. Along with the expected intervalley scattering oscillations with a wavelength of $\lambda_{\Delta K}= 2\pi/\Delta K\simeq3.7\,\mbox{\AA}$ (${\bf \Delta K= K'-\bf K}$ connects two adjacent Dirac points), the filtered images present a couple of dislocations in the vicinity of the H adatom. Trained eyes can track them in raw images (Fig~\ref{STM Dislocations}e and supplementary information). STM imaging after manipulation of the H atom\cite{Gonzalez-Herrero:2016aa} never revealed structural defects in graphene proving that the dislocations appear only in Friedel oscillations. These dislocations allow to measure the Berry phase since they are real-space consequences of graphene's pseudospin winding around a Dirac cone as we shall now show.

\begin{figure*}[t!]
	\centering
	\includegraphics[width=0.9\textwidth]{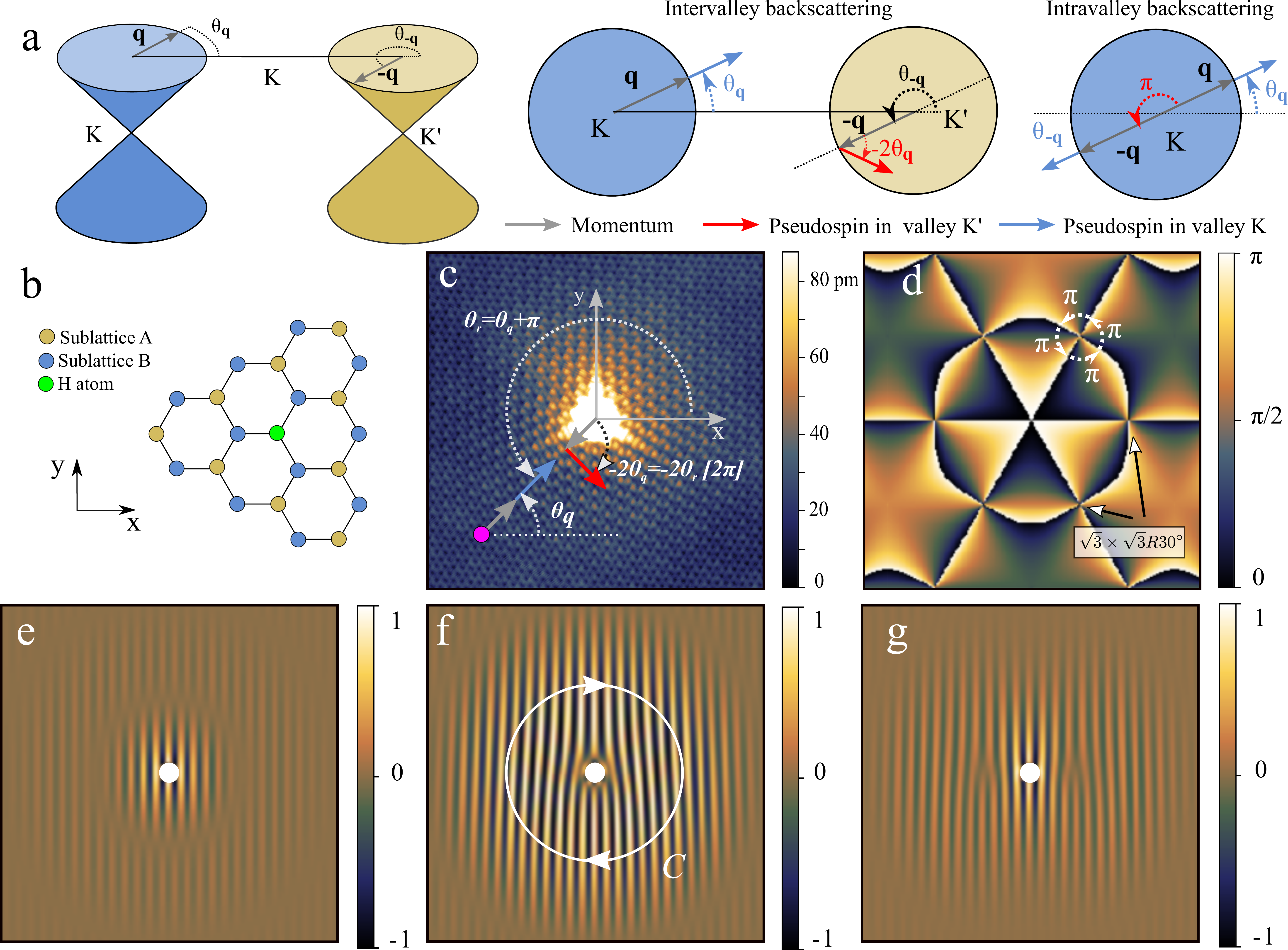}
	\caption{\small \textbf{Theoretical description of the dislocations in Friedel oscillations.} 
		{\bf a}, Backscattering process in graphene. Intervalley backscattering between wavevector states ${\bf q}$ and ${-\bf q}$ belonging to nearest-neighbour valleys ${\bf K}$ and ${\bf K'}$ leads to a rotation of the pseudo spin of $-2\theta_{\bf q}$. Intravalley backscattering rotates the pseudospin by $\pi$. 
		{\bf b}, Honeycomb lattice of graphene with a chemisorbed H adatom on sublattice A. {\bf c}, Relation between the STM tip position and the pseudospin rotation in intervalley backscattering by a H atom. The STM image is the same as in Fig.~\ref{STM Dislocations}a but was rotated for direct comparison with the theory. The STM tip is represented by the purple dot. {\bf d}, Phase image of the Fourier transform of the theoretical density modulation $\rho_{A+B}({\bf r})$ (to compare with Fig.~\ref{STM Dislocations}c). The image is 59 nm$^{-1}$$\times$ 59 nm$^{-1}$. {\bf e, f, g}, Calculated charge density modulation induced by intervalley scattering on the A sublattice ({\bf e}) on the B sublattice ({\bf f}) and total charge density modulation ({\bf g}). The modulations have been normalized to 1. The images are 10\,nm $\times$ 10\,nm and the theory is integrated from 0 to 0.4\,eV, as in the STM experiments of Fig.\,\ref{STM Dislocations}. The white disk depicts the H adatom.}
	\label{Scattering}
\end{figure*}

Friedel oscillations in STM images are dominated by backscattering processes along iso-energy contours.\cite{Sprunger:1997aa} At a given tip position, the amplitude of the Friedel oscillation probed by the STM is governed by the interference of the electronic wave pointing toward the H atom placed at ${\bf r}=0$ and its reflection from the adatom. As a consequence,  the angle $\theta_{\bf q}$ parametrizing the momentum ${\bf q}$ of the incident electron is directly related to the angle $\theta_{\bf r}=\theta_{\bf q}+\pi$ indexing the tip position (see Fig.~\ref{Scattering}a and \ref{Scattering}c where the angles are defined with respect to the direction ${\bf \Delta K}$). In graphene, $\theta_{\bf q}$ also defines the momentum-locked pseudo-spin of the incident electronic wave in valley {\bf K}. Intravalley back-scattering involves a rotation of the pseudospin that is always $\pi$ so that the interference is destructive at the leading order  (Fig.~\ref{Scattering}a).\cite{Cheianov:2006aa,Brihuega:2008,Dutreix:2016aa} In contrast, back-scattering from valley {\bf K} to valley {\bf K'} involves a rotation of the pseudospin by the angle $-2\theta_{\bf q}=-2\theta_{\bf r}~[2\pi]$ (cf. Fig.~\ref{Scattering}a, Fig.\,\ref{Scattering}c) which does not kill the associated Friedel oscillation but leads to the peculiar interference pattern we observe. This pattern is linked to graphene's Berry phase because circling the STM tip around the impurity implies circling the incident electron's ${\bf q}$ on a closed iso-energy contour around the Dirac point in reciprocal space as $\theta_{\bf q}$ is locked on $\theta_{\bf r}$ (supplementary video). This is analogous to the trajectory of the momentum on a cyclotron orbit but the movement of the STM tip replaces the adiabatic transport of electrons in magneto-transport measurements.

More formally, an isolated H adatom constitutes an atomic scatterer which induces both intra- and inter-valley scattering. It may be modeled by a Dirac delta potential $V_{0}\,\delta(\bf r)$ ($V_{0}\gg$1\,eV) \cite{katsnelson2012graphene}. Elastic scattering of Dirac electrons on such potential has an analytical solution that is non-perturbative in $V_{0}$ (Supplementary information). For an adatom located on sublattice A (Fig.~\ref{Scattering}b) and for a given direction of intervalley scattering, it yields a modulation of the charge density around the adatom that reads:

\begin{align}\label{Eq : Intervalley Density}
\delta\rho({\bf \Delta K},{\bf r}, V_{b})
=&\delta\rho_{A}(r, V_{b})\cos({\bf \Delta K} \cdot {\bf r})\notag \\
+\xi\xi'&\delta\rho_{B}(r, V_{b})\cos({\bf \Delta K} \cdot {\bf r}-(\xi-\xi')\theta_{\bf r}) ~.
\end{align}

Here, the two terms correspond to the density modulation on the A and B sublattices respectively.
For intravalley scattering (${\bf \Delta K}=0$ and the valley index $\xi'=\xi=1$), the charge modulation is defined entirely by $\delta\rho_{A}$ and $\delta\rho_{B}$ which describe the universal static response of conduction electrons to impurities, i.e, the usual Friedel oscillations.\cite{Friedel:1952aa} They allow to determine the spectral properties via their $2q_{F}$-wave-vector dependence\cite{Rutter:2007}and have an unconventional decay in graphene because of the $\pi$ rotation of the pseudospin in intravalley backscattering.\cite{Cheianov:2006aa,Brihuega:2008,Dutreix:2016aa}  Their expressions are given in Supplementary information. These oscillations have a long period $\lambda_{F}/2=\pi/q_{F}\simeq5.2$\,nm for $q_{F}$ fixed by the experimental tunneling bias $V_{b}=0.4$\,V.

Extra oscillations appear in Eqs.~\ref{Eq : Intervalley Density} for intervalley scattering (${\bf \Delta K}\neq0$, $\xi'=-\xi$). Contrary to usual Friedel oscillations, their wave-length $\lambda_{\Delta K}$, is independent of energy so that they are not smeared by the integration on the STM bias window (See the experimental proof from $dI/dV$ maps in the supplements). The corresponding modulation of electronic density is plotted in Fig.\,\ref{Scattering}e, f and g for a given direction of intervalley scattering. Importantly, the angle $-2\theta_{\bf r}$, which turns out to be the real space representation of the pseudospin rotation in intervalley backscattering (supplementary information), appears as an additional phase shift in the density modulation on sub-lattice B. It encodes all the {\bf q} dependence of the pseudospin and maps its singularity at the Dirac cone apex into a singularity in the real space from which the wave front dislocation emerges.  

Topological defects in waves were introduced by Nye and Berry who showed that the dislocations in the radio-echos sounding the ice sheet of Antartica resulted from phase singularities in a complex scalar field describing the wave propagation.\cite{Nye:1974aa} Such topological defects in waves are ubiquitous in physics from fluids\cite{Berry_1980,Berry:2000aa} to singular optics\cite{DENNIS2009293,Rafayelyan:16} and condensed matter. \cite{kosterlitz1973ordering,Feynman:1955aa,Abrikosov:1957aa}
We follow Nye and Berry in defining the complex scalar field $\varrho_B({\bf r}) = |\varrho_B({\bf r})|\,e^{i\varphi_B({\bf r})}$, whose real part describes the Friedel oscillation on the B sublattice (the second term in Eq.\,\ref{Eq : Intervalley Density}). The phase $\varphi_B({\bf r})={\bf \Delta K}\cdot{\bf r}-2\theta_{\bf r}$ is singular at ${\bf r}=0$. It can be regarded as a potential which gradient is the sum of a uniform field and a vortex.~\cite{Nye:1974aa} In our case, the uniform field represents the standing electronic wave associated to intervalley back-scattering and the vortex, its perturbation by the pseudospin rotation. The circulation of this field is the phase accumulated along a closed path $C$. It is necessarily quantized to an integer topological number $N$ of $2\pi$ because $\varrho_B({\bf r})$ is a single-valued field and must return to the same observable electron density after circulating along the closed circuit. In singular optics $N$ is called the {\it charge} of the phase singularity. It represents the number of additional wave fronts necessary to accommodate for the phase accumulated along $C$. It is obviously 0 if the closed path does not enclose the phase singularity. For path enclosing the singularity (Fig.~\ref{Scattering}f), the gradient circulation of $\varphi_B({\bf r})$ is equal to the winding of $-2\theta_{\bf r}$ and hence of that of $-2\theta_{\bf q}$. Since the Berry phase in graphene $\gamma=\pi$ is given by half the winding of $\theta_{\bf q}$ (supplementary information) it follows that $2N\pi=4\gamma$ for a clockwise oriented contour. The $N=2$ additional wavefronts seen in Fig.~\ref{Scattering}f are therefore signature of the Berry phase in graphene and prove the existence of Dirac cones.  
We note that given the quality of the STM image, this winding of $4\gamma=4\pi$ can also be directly retrieved from the phase of the Fourier transform~\cite{Dutreix:2016aa} as shown in Figs~\ref{STM Dislocations}c and \ref{Scattering}d.

The contribution from sublattice A to the total electron density modulation only alters the shape of the dislocation which is a robust topological feature (See the discussion of the total scalar field $\varrho_{A+B}({\bf r})$ in Supplementary information). The dislocations are shifted from $r=0$ in the direction ${\bf \Delta K}$ in agreement with experiments (Fig.~\ref{STM Dislocations}d and e and Fig.~\ref{Scattering}g).

\begin{figure*}[th!]
	\centering
	$\begin{array}{cc}
	\includegraphics[width=0.9\textwidth]{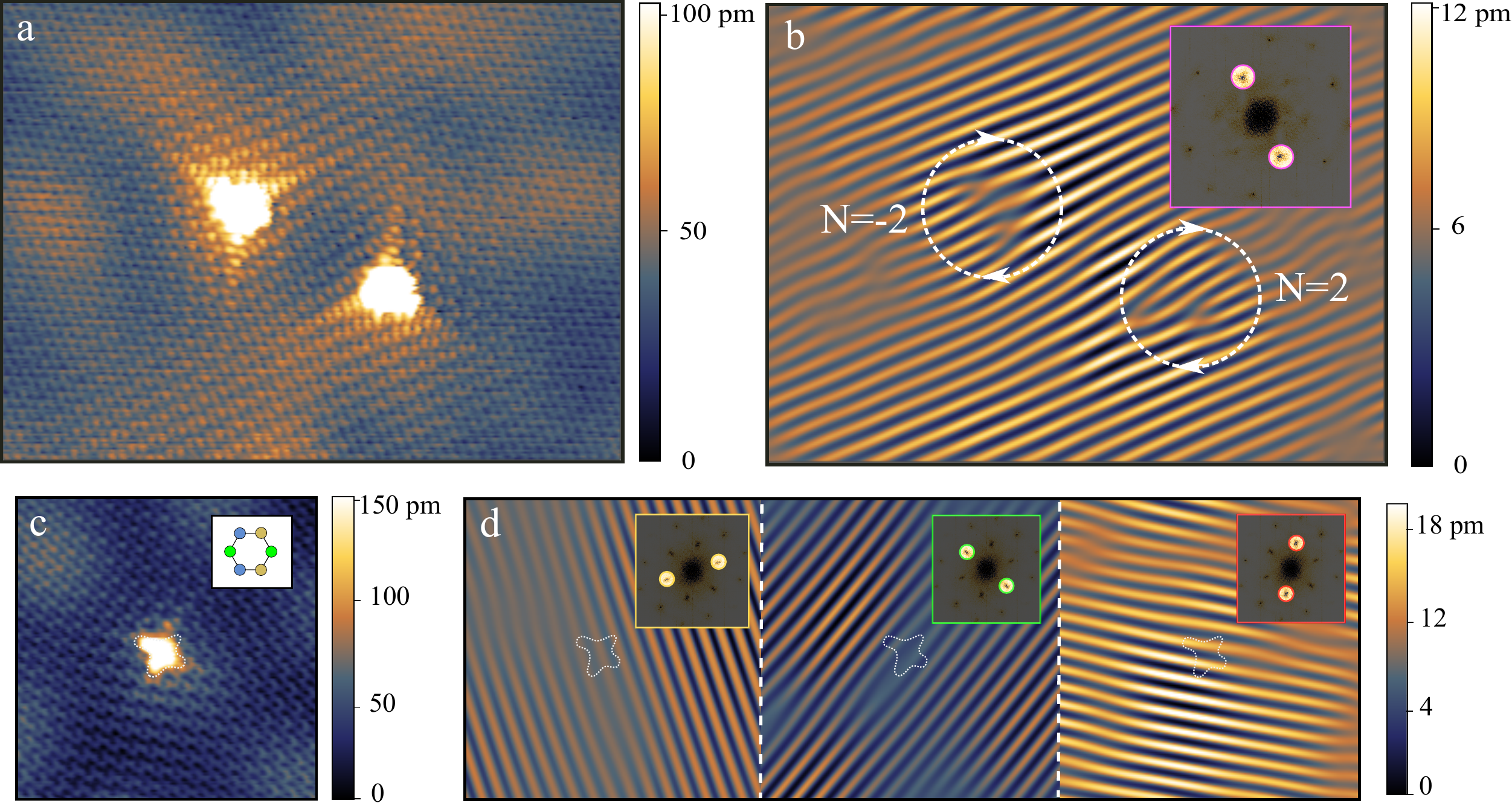}
	\end{array}$
	\caption{\small \textbf{Friedel oscillation around adatoms situated on different sublattices.} {\bf a}, STM image of the static interferences around two hydrogen adatoms chemisorbed on different sublattices of graphene. The image is 12 nm $\times$ 9 nm in size. The tunneling bias is $V_b=0.4$~V and the tunneling current is $i_t=33.1$~pA. {\bf b}, FFT filtered image of the image in panel a). The inset shows the filter used. The sign of the topological charge of each phase singularity depends on the sublattice the adatoms belong to.  {\bf c}, Topography STM image ($6.8\times 6.8$\,nm$^{2}$) of a H-adatom dimer chemisorbed on neighbourg A and B sites (Schematic illustration in the inset which has the same color code as Fig.~\ref{Scattering}b). The tunnelling bias is $V_b=-0.4$~V and the tunnelling current is $i_t=51.5$~pA. {\bf d}, The image of panel c is FFT filtered along the 3 directions of intervalley scattering. In panels {\bf b} and {\bf d} the inset are 11 nm$^{-1}\times$11 nm$^{-1}$.}
	\label{STM dimmer}
\end{figure*}

The H atom can be placed on different sublattice as inferred from the different orientation of the tripod shape of the H signal in the image of Fig~\ref{STM dimmer}a.~\cite{Gonzalez-Herrero:2016aa} For a given orientation of closed path around the impurity, the sign of $N$ is opposite for the two cases (Fig.~\ref{STM dimmer}b). This is because the two configurations relate to one another via inversion symmetry with respect to the center of a $C$-$C$ bound. Since the underlying lattice of graphene is bipartite, this further means that this single-particle topological signature of the sublattice imbalance also relates, via Lieb's theorem \cite{Lieb1989aa}, to the spontaneous magnetic moments induced by the electron interactions at half filling. \cite{Gonzalez-Herrero:2016aa}. Contrary to this many-body effect, the dislocations in Friedel oscillations are independent of doping (Supplementary information). Figure~\ref{STM dimmer}c and~\ref{STM dimmer}d show that if two H atoms are placed on neighbouring carbon atoms there is no dislocation in the intervalley scattering signal. This  results from the annihilation of dislocations of opposite $N$ and illustrates that disorder has to brake sublattice symmetry (see supplementary information). 

In quantum mechanics, wavefront dislocations have been predicted for scalar wavefunctions such as the Aharonov–Bohm wavefunction, but were thought to be unobservable owing to the U(1) gauge invariance of the density\cite{Berry_1980}. We have demonstrated that dislocations appear in the charge density of vectorial wavefunctions, the components of
which can interfere with each other by scattering between distant time-reversed valleys.
Since wavefront dislocations arise from phase singularities \cite{Nye:1974aa} which relate to the topological properties of band structures for vectorial wave functions, wavefront dislocations in Friedel oscillations can lead to the identification of relativistic and topological phases as already established theoretically for rhombohedral graphite\cite{Dutreix:2016aa} and 1D insulator\cite{Dutreix:2017aa}. This method of determining the topological properties of band structures is complementary to transport measurements under strong magnetic field. However, contrary to transport measurements where it poisons quantum Hall measurements, disorder turns out to be here an asset as long as an area of few tens of nm$^2$ with a point like scatterer is available on the surface.

\section*{Methods}
\subsection*{Sample preparation}
Graphene was grown on 6H-SiC[000$\bar{1}$] by thermal annealing following the recipe described in Ref.~\onlinecite{Varchon2008}. This leads to the growth of graphene layers whose low energy physics is that of single layer graphene owing to the decoupling by rotationnal disorder.\cite{Haas2008} The doping of the top graphene layer could be controlled by the number of underlying layers which is governed by the annealing temperature/time.\cite{Gonzalez-Herrero:2016aa} The results presented on the main manuscript were obtained on a thick multilayer (more than 5 graphene layers) in which the subtrate is too far away to dope significantly the layers by charge transfer (see also supplementary information). A thinner one (2-4 graphene layers) was prepared to investigate the effect of doping (see supplementary information).
Hydrogen atoms were deposited on the surface of graphene on SiC by thermal dissociation of H$_2$ in a home made hydrogen atom beam source in UHV conditions.\cite{Hornekaer2006} A molecular $H_2$ beam is passed through a hot W filament held at 1900K. The pristine graphene substrate is placed 10 cm away from the filament, held at RT during atomic H deposition and subsequently cooled down to 5K, the temperature at which we carried out all STM/STS experiments presented here. $H_2$ pressure is regulated by a leak valve and fixed to 3$\times10^{-7}$ torr as measured in the preparation chamber for the present experiments. The atomic H coverage was adjusted by changing the deposition times between 200-60s which corresponded to final coverages between 0.10-0.03 H atoms/nm$^2$

\subsection*{STM measurements}

The STM measurements were performed {\it in situ} using a home made low temperature scanning tunneling microscope operating at 5~K in ultrahigh vacuum. Images presented in the main manuscript were performed in constant current mode. Conductance spectra and images presented in the supplements were taken using a lock-in technique, with an ac voltage (frequency: 830 Hz, amplitude: 1-2 mV rms) added to the dc sample bias. 

\bibliographystyle{nature}
\bibliography{references}

\section*{Acknowledgements}

The authors thank P. Mallet, J-Y. Veuillen and JM. G\'{o}mez Rodriguez for experimental support.
HG-H and IB acknowladge support from MINECO (grants MAT2016-80907-P and PCIN-2015-030) and Fundación Ramón Areces. MIK acknowledges a support of NWO via Spinoza Prize.

\section*{Author contributions}
HG-H and IB performed the experiments. VTR discovered the dislocations which were explained with the theory derived by CD. CD and VTR wrote the manuscript with the input of all authors. VTR coordinated the collaboration.

\section*{Competing financial interests}
The authors declare no competing financial interests.

\section*{Data availability}
The datasets generated during and/or analysed during the current study are available from the corresponding author on reasonable request.

\end{document}